\documentclass[runningheads]{llncs}

\usepackage{amssymb}
\usepackage{graphicx}

\usepackage{url}

\newcommand\nc{\newcommand}

\newtheorem{fact}[theorem]{Fact}

\def\boxit#1{\vbox{\hrule\hbox{\vrule\kern4pt
  \vbox{\kern1pt#1\kern1pt}
\kern2pt\vrule}\hrule}}

\nc{\crl}[2]{\begin{corollary}\label{crl:#1} #2 \end{corollary}}
\nc{\dfn}[2]{\begin{definition}\label{def:#1} #2 \end{definition}}
\nc{\lem}[2]{\begin{lemma}\label{lem:#1} #2 \end{lemma}}
\nc{\prp}[2]{\begin{proposition}\label{prp:#1} #2
\end{proposition}}
\nc{\thm}[2]{\begin{theorem}\label{thm:#1} #2\end{theorem}}
\nc{\fac}[2]{\begin{fact}\label{fact:#1} #2 \end{fact}}

\nc{\eqn}[2]{\begin{eqnarray}\label{eqn:#1} #2 \end{eqnarray}}

\nc{\fig}[4]{\begin{figure}[h]
\begin{center}
\includegraphics[width=#2\textwidth]{#4}
\end{center}
\caption{#3}\label{fig:#1}
\end{figure}}

\nc{\tbl}[3]{\begin{table}[hbt] #3 \caption{#2} \label{tab:#1}
\end{table}}

\nc{\refc}[1]{Corollary~\ref{crl:#1}}
\nc{\refd}[1]{Definition~\ref{def:#1}}
\nc{\reff}[1]{Figure~\ref{fig:#1}}
\nc{\refl}[1]{Lemma~\ref{lem:#1}}
\nc{\refp}[1]{Proposition~\ref{prp:#1}}
\nc{\reft}[1]{Theorem~\ref{thm:#1}}
\nc{\refe}[1]{(\ref{eqn:#1})}
\nc{\reftb}[1]{Table~\ref{tab:#1}}

\nc{\reffc}[1]{Fact~\ref{fact:#1}}

\begin{document}



\title{New Branching Rules: Improvements on Independent Set and Vertex Cover in Sparse Graphs\footnote{The paper was presented at the 2nd
annual meeting of asian association for algorithms and computation
(AAAC 2009), April 11-12, 2009, Hangzhou, China.}}

\titlerunning{Independent Set and Vertex Cover in Sparse graphs}

%
%
\author{Mingyu Xiao}
\authorrunning{}


\institute{School of Computer Science and Engineering\\
University of Electronic Science and Technology of China \\
Chengdu 610054, CHINA\\
Email: {\tt myxiao@gmail.com}
}
%
%

\toctitle{Maximum Independent Set} \tocauthor{} \maketitle

\begin{abstract}
We present an $O^*(1.0919^n)$-time algorithm for finding a maximum
independent set in an $n$-vertex graph with degree bounded by $3$,
which improves the previously known algorithm of running time
$O^*(1.0977^n)$ by Bourgeois, Escoffier and Paschos [IWPEC 2008].
We also present an $O^*(1.1923^k)$-time algorithm to decide if a
graph with degree bounded by $3$ has a vertex cover of size $k$,
which improves the previously known algorithm of running time
$O^*(1.1939^k)$ by Chen, Kanj and Xia [ISAAC 2003].

Two new branching techniques, \emph{branching on a bottle} and
\emph{branching on a $4$-cycle}, are introduced, which help us to
design simple and fast algorithms for the maximum independent set
and minimum vertex cover problems and avoid tedious branching
rules.

\vspace{5mm}\noindent {\bf Key words.} \ \ Graph Algorithm,
Independent Set, Vertex Cover, Sparse Graph
\end{abstract}

\section{Introduction}\label{sec_intr}
The \emph{maximum independent set} problem (MIS), to find a
maximum set of vertices in a graph such that there is no edge
between any two vertices in the set, is one of the basic NP-hard
optimization problems and has been well studied in the literature,
in particular in the line of research on worst-case analysis of
algorithms for NP-hard optimization problems. In 1977, Tarjan and
Trojanowski~\cite{Tarjan:IS} published the first algorithm for
this problem, which runs in $O^*(2^{n/3})$ time and polynomial
space. Later, the running time was improved to $O^*(2^{0.304n})$
by Jian~\cite{Jian:Is}. Robson~\cite{Robson:IS} obtained an
$O^*(2^{0.296n})$-time polynomial-space algorithm and an
$O^*(2^{0.276n})$-time exponential-space algorithm. In a technical
report~\cite{Robson:IS_1}, Robson also claimed better running
times. Recently, Fomin et al.~\cite{Fomin:is} got a simple
$O^*(2^{0.288n})$-time polynomial-space algorithm by using the
``Measure and Conquer" method. There is also a considerable amount
of contributions to the maximum independent set problem in sparse
graphs, especially in degree-$3$
graphs~\cite{Beigel:is},\cite{Chen:labeled3vc},\cite{xiao:IS3},\cite{Bourgeois:3IS}.
We summarize the results on low-degree graphs as well as general
graphs in Table~\ref{table1}.

\begin{table}[htp]\label{table1}
\begin{tabular}{|l|l|l|l|}\hline
\textbf{Authors} & \textbf{Running times} &\textbf{References}& \textbf{Notes}\\
\hline \hline Tarjan \& Trojanowski &$O^*(1.2600^n)$ for MIS&
1977~\cite{Tarjan:IS}&$n$: number of vertices\\ \hline Jian & $O^*(1.2346^n)$ for MIS& 1986~\cite{Jian:Is}&\\
\hline Robson &$O^*(1.2109^n)$ for MIS &1986~\cite{Robson:IS}&Exponential space\\
\hline
 Beigel & $O^*(1.0823^m)$ for MIS
& 1999~\cite{Beigel:is}&$m$: number of edges\\
&$O^*(1.1259^n)$ for $3$-MIS & &$3$-MIS: MIS in degree-$3$ graphs\\
 \hline Robson &$O^*(1.1893^n)$ for MIS &2001~\cite{Robson:IS_1}&Partially computer-generated\\
 \hline Chen et al. & $O^*(1.1254^n)$ for $3$-MIS & 2003~\cite{Chen:labeled3vc}&\\
\hline Xiao et al. & $O^*(1.1034^n)$ for $3$-MIS & 2005~\cite{xiao:IS3}&Published in Chinese\\
\hline Fomin et al. & $O^*(1.2210^n)$ for MIS & 2006~\cite{Fomin:is}&\\
\hline Fomin \& H\o ie & $O^*(1.1225^n)$ for $3$-MIS & 2006~\cite{Fomin:cubicgraph}&\\
\hline F{\"u}rer & $O^*(1.1120^n)$ for $3$-MIS & 2006~\cite{Furer:ISsparse}&\\
\hline Razgon & $O^*(1.1034^n)$ for $3$-MIS & 2006~\cite{Razgon:3IS}&\\
\hline Bourgeois et al. & $O^*(1.0977^n)$ for $3$-MIS &
2008~\cite{Bourgeois:3IS} &\\
\hline Xiao &$O^*(1.0919^n)$ for $3$-MIS& This paper & \\
\hline
\end{tabular}
\vspace{3mm} \caption{Exact algorithms for the maximum independent
set problem}
\end{table}
In the literature, there are several methods of designing
algorithms for finding maximum independent sets in graphs. One
method is to find a minimum vertex cover (a set of vertices such
that each edge in the graph has at least one endpoint in the set),
and then to get a maximum independent set by taking all the
remaining vertices, such as the algorithms presented
in~\cite{Chen:labeled3vc},\cite{VC2005}. In this kind of
algorithms, the dominating part of the running time is the running
time for finding a minimum vertex cover. Another method is based
on the search tree method. We will use a branch-and-reduce
paradigm. We choose a parameter, such as the number of vertices or
edges or others, as a measure of the size of the problem. When the
parameter is zero or a negative number, the problem can be solved
in polynomial time. We branch on the current graph $G$ into serval
graphs $G_1$, $G_2$, $\cdots, G_l$ such that the parameter $r_i$
of graph $G_i$ is less than the parameter $r$ of graph $G$
($i=1,2,\cdots, l$), and a maximum independent set in $G$ can be
found in polynomial time if a maximum independent set in each of
the $l$ graphs $G_1$, $G_2$, $\cdots, G_l$ is known. By this
method, we can build up a search tree, and the exponential part of
the running time of the algorithm is corresponding to the size of
the search tree. The running time analysis leads to a linear
recurrence for each node in the search tree that can be solved by
using standard techniques. Let $C(r)$ denote the worst-case size
of the search tree when the parameter of graph $G$ is $r$, then we
get recurrence relation $C(r)\leq \sum_{i=1}^l C(r_i)$. Solving
the recurrence, we get $C(r)=[\alpha(r, r_1, r_2, \cdots,
r_l)]^r$, where $\alpha(r, r_1, r_2, \cdots, r_l)$ is the largest
root of the function $f(x)=1-\sum_{i=1}^l x^{r_i-r}$. As for the
measure (the parameter $r$), a natural one is the number of
vertices or edges in the graph. Most previous algorithms for the
maximum independent set problem are analyzed by using the number
of vertices as a
measure~\cite{Tarjan:IS},\cite{Jian:Is},\cite{Robson:IS},\cite{Fomin:is}.
The number of edges is considered in Beigel's
algorithm~\cite{Beigel:is}. There are also some other measures.
Xiao et al.~\cite{xiao:IS3} used the number of degree-$3$ vertices
as a measure to analyze algorithms and got an $O^*(1.1034^n)$-time
algorithm for MIS in degree-$3$ graphs. Unfortunately, that paper
was published in Chinese. Recently, Razgon~\cite{Razgon:3IS} also
got an $O^*(1.1034^n)$-time algorithm for MIS in degree-$3$ graphs
by measuring the number of degree-$3$ vertices. But the two
algorithms are totally different. F{\"u}rer~\cite{Furer:ISsparse}
designed an algorithm for MIS in degree-$3$ graphs by tackling
$m-n$, where $m$ is the number of edges and $n$ the number of
vertices. Based upon a refined branching with respect to
F{\"u}rer's algorithm, Bourgeois et al.~\cite{Bourgeois:3IS} got
the current best algorithm for MIS in degree-$3$ graphs with
running time $O^*(1.0977^n)$. In this paper, we still use the
number of degree-$3$ vertices as a measure to analyze our
algorithm. Based on two new branching rules, \emph{branching on a
bottle} and \emph{branching on a $4$-cycle}, we design an even
faster algorithm for MIS in degree-$3$ graphs, which runs in
$O^*(1.0919^n)$ time. Our algorithm is simple and does not contain
many branching rules. Furthermore, it can be used to solve the
\emph{$k$-vertex cover} problem (to decide if the graph has a
vertex cover of size $k$) in degree-$3$ graphs in $O^*(1.1923^k)$
time, which improves the previously known result of
$O^*(1.1939^k)$ by Chen et al.~\cite{Chen:labeled3vc}.

\section{Preliminaries}
We shall try to be consistent in using the following notation. The
number of vertices in a graph will be denoted by $n$ and the
number of degree-$3$ vertices (vertices of degree $\geq 4$ will
also be counted with a weight) by $r$. For a vertex $v$ in a
graph, $d(v)$ is the degree of $v$, $N(v)$ the set of all
neighbors of $v$, $N[v]=N(v)\cup\{v\}$ the set of vertices with
distance at most $1$ from $v$, and $N_2(v)$ the set of vertices
with distance exactly $2$ from $v$. We say edge $e$ is incident on
a vertex set $V'$, if at least one endpoint of $e$ is in $V'$. In
our algorithm, when we remove a set of vertices, we also remove
all the edges that are incident on it. Throughout the paper we use
a modified $O$ notation that suppresses all polynomially bounded
factors. For two functions $f$ and $g$, we write $f(n) =
O^*(g(n))$ if $f(n) = O(g(n)poly(n))$, where $poly(n)$ is a
polynomial.

Our algorithms are based on the branch-and-reduce paradigm. We
will first apply some reduction rules to reduce the size of
instances of the problem. Then we apply some branching rules to
branch on the graph by including some vertices in the independent
set or excluding some vertices from the independent set. In each
branch, we will get a maximum independent set problem in a graph
with a smaller measure. Next, we introduce the reduction rules and
branching rules that will be used in our algorithms.

\subsection{Reduction Rules}
There are several standard preprocesses to reduce the size of
instances of the problem. \emph{Folding a degree-$1$ or degree-$2$
vertex} and \emph{removing a dominated vertex} are frequently used
rules. Besides these reduction rules, we still need to reduce some
other local structures called \emph{$2$-$3$ structure},
\emph{$3$-$3$ structure} and \emph{$3$-$4$ structure}.

\vspace{2mm}\noindent \textbf{Folding a degree-1 vertex}\\
\emph{Folding a degree-1 vertex $v$ means removing $v$ and $u$
from the graph, where $u$ is the unique neighbor of $v$.}

\vspace{2mm}\noindent \textbf{Folding a degree-2 vertex}\\
\emph{Folding a degree-$2$ vertex $v$ (with two neighbors $u$ and
$w$) means\\
$(a)$ removing $v$, $u$ and $w$ from the
graph, when $u$ and $w$ are adjacent.\\
$(b)$ removing $v$, $u$ and $w$ from the graph and introducing a
new vertex $s$ that is adjacent to all neighbors of $u$ and $w$ in
$G$ $($except the removed vertex $v)$, when $u$ and $w$ are
nonadjacent.}

\vspace{0mm}Please refer to \reff{folding} for an illustration of
the operation in case $(b)$ of folding a degree-$2$ vertex. Let
$\alpha(G)$ denote the size of a maximum independent set of graph
$G$ and $G^\star(v)$ the graph after folding a degree-$1$ or
degree-$2$ vertex $v$. Then we have the following
lemma.\lem{fold_1}{For any degree-$1$ or degree-$2$ vertex $v$ in
graph $G$, \vspace{-2mm}$$\alpha(G)=1+\alpha(G^\star(v)).$$ }
\vspace{-4mm}\fig{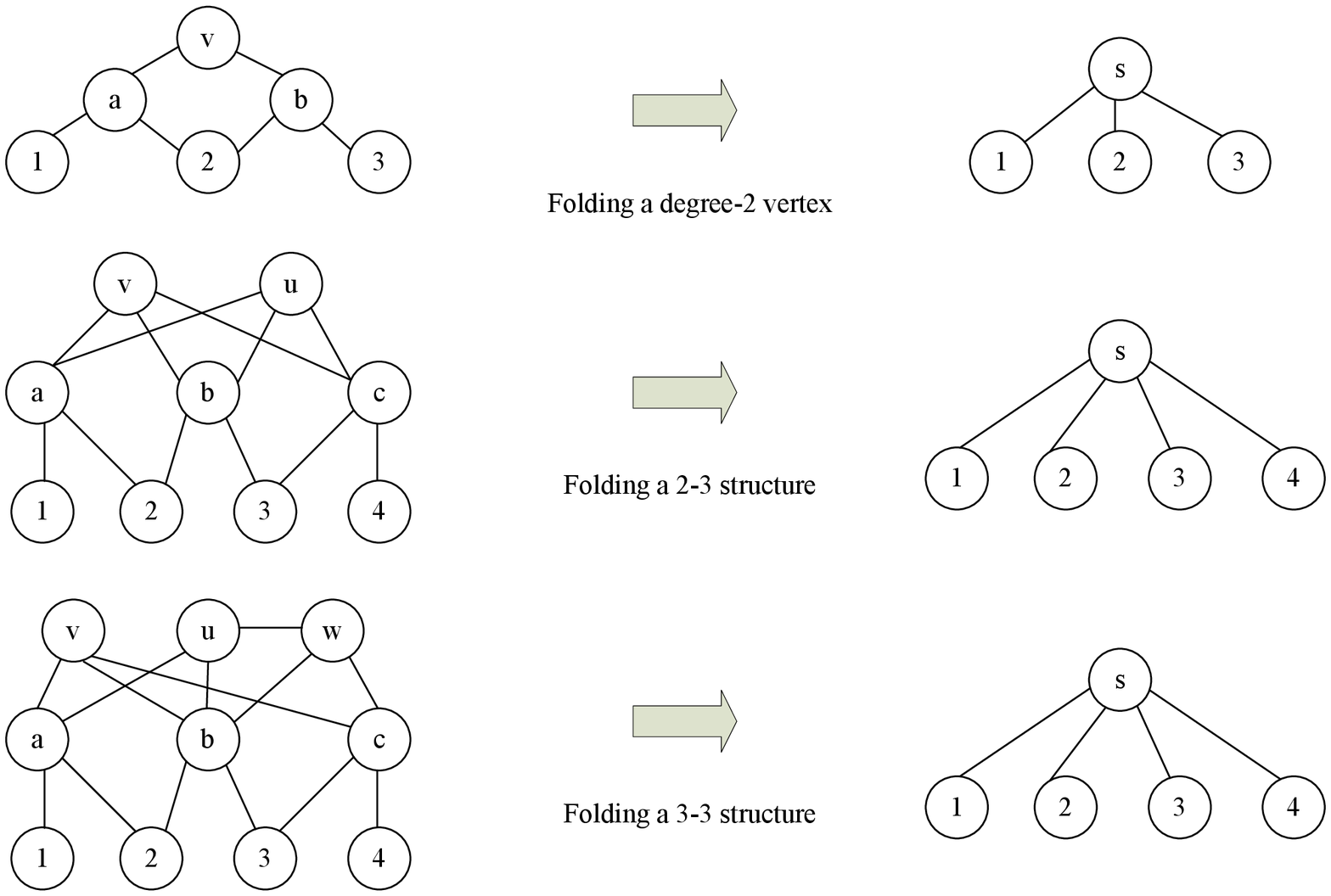}{1}{Illustrations of folding
operations}{folding}

The correctness of folding a degree-$1$ or degree-$2$ vertex has
been discussed in many pervious papers. In fact, general folding
rules are known in the literature, which can deal with a vertex of
degree $\geq3$ or a set of independent
vertices~\cite{VC2005},\cite{Fomin:is}. In this paper, we still
need to fold the following three local structures called $2$-$3$
structure, $3$-$3$ structure and $3$-$4$ structure.

Let $u$ and $v$ be two independent degree-$3$ vertices, if they
have three common neighbors $a,b$ and $c$, then we say that the
five vertices compose a \emph{$2$-$3$ structure} (see
\reff{folding}), and denote it by $\{u,v\}$-$\{a,b,c\}$. Let $v$
be a degree-$3$ vertex, and $u$ and $w$ two adjacent vertices of
degree $\geq3$. If $N(u)\cup N(w)-\{u,w\}= N(v)$, then we say that
the six vertices $\{u, v, w\}\cup N(v)$ compose a \emph{$3$-$3$
structure} (see \reff{folding}), and denote it by
$\{v,u,w\}$-$\{a,b,c\}$. Let $u, v$ and $w$ be three independent
vertices of degree $\geq3$ , if they have exact four neighbors
$a,b,c$ and $d$, then we say that the seven vertices compose a
\emph{$3$-$4$ structure}, and denote it by
$\{u,v,w\}$-$\{a,b,c,d\}$.

\vspace{2mm}\noindent \textbf{Folding a $2$-$3$ structure, $3$-$3$ structure or $3$-$4$ structure}\\
\emph{Let $A$-$B$ be a $2$-$3$ structure or $3$-$3$ structure or
$3$-$4$ structure. Folding $A$-$B$ means\\
$(a)$ removing $A\cup B$ from the
graph, when $B$ is not an independent set.\\
$(b)$ removing $A\cup B$ from the graph and introducing a new
vertex $s$ that is adjacent to all neighbors of vertices in $B$
$($except the removed vertices$)$, when $B$ is an independent
set.}

\lem{fold_2}{If graph $G$ has a $2$-$3$ structure or $3$-$3$
structure, then
$$\alpha(G)=2+\alpha(G^\star_2),$$ where $G^\star_2$ is the graph
after folding a $2$-$3$ structure or $3$-$3$ structure in $G$.

If graph $G$ has a $3$-$4$ structure, then
$$\alpha(G)=3+\alpha(G^\star_3),$$ where $G^\star_3$ is the graph
after folding a $3$-$4$ structure in $G$. }

A degree-$2$ vertex can be regarded as a $1$-$2$ structure
according to our definitions. In fact, a degree-$2$ vertex,
$2$-$3$ structure and $3$-$4$ structure are special cases
described in Lemma $2.4$ in \cite{VC2005}. The $3$-$3$ structure
is for the first time being introduced. The correctness of folding
an $A$-$B$ structure (a $1$-$2$ structure, $2$-$3$ structure,
$3$-$3$ structure or $3$-$4$ structure) follows from this
observation: When $B$ is not an independent set, there is a
maximum independent set that contains $A$ (or two independent
vertices in $A$, when $A$-$B$ is a $3$-$3$ structure). When $B$ is
an independent set, there is a maximum independent set that
contains either $B$ or $A$ (or two independent vertices in $A$,
when $A$-$B$ is a $3$-$3$ structure). We ignore the detailed proof
here.

\vspace{2mm}\noindent \textbf{Dominance}\\
\emph{If there are two vertices $v$ and $u$ such that
$N[u]\subseteq N[v]$, we say $u$ dominates $v$.} \lem{}{If vertex
$v$ is dominated by any other vertex in graph $G$, then
\vspace{-2mm}$$\alpha(G)=\alpha(G-\{v\}).$$ }

\dfn{}{A graph is called a \emph{reduced graph}, if it has no
degree-$1$ vertex, degree-$2$ vertex, dominated vertex, $2$-$3$
structure, $3$-$3$ structure or $3$-$4$ structure.}

\subsection{Branching Rules}

Next we introduce two branching techniques, \emph{branching on a
bottle} and \emph{branching on a $4$-cycle}, which are simple and
obvious, but can  avoid tedious branching rules in the description
of the algorithms.

Let $a$ be a degree-$3$ vertex, and $b,c,d$ the three neighbors of
$a$. If two neighbors of $a$, say $c$ and $d$, are adjacent, then
we say that the four vertices compose a \emph{bottle} and denote
it by $b$-$a$-$\{c,d\}$.

\lem{bottlebranch}{Let $b$-$a$-$\{c,d\}$ be a bottle in graph $G$,
then there is a maximum independent set $S$ in $G$ such that
either $a\in S$ or $b\in S$.} \proof{If $b$ is not in a maximum
independent set, we can directly remove $b$ from the graph. In the
remaining graph $a$ becomes a degree-$2$ vertex and the two
neighbors of it are adjacent. In this case, there is a maximum
independent set that contains $a$.}

Based on \refl{bottlebranch}, we get the following branching rule.

\vspace{2mm}\noindent \textbf{Branching on a bottle}\\
\emph{Branching on a bottle $b$-$a$-$\{c,d\}$ means branching by
either including $a$ in the independent set or including $b$ in
the independent set.}

\noindent\textbf{Note. }In fact, we can fold a bottle by using the
general folding rule mentioned in~\cite{Fomin:is} (also
in~\cite{Beigel:is}), but this folding rule is helpless for our
analysis, especially when the three neighbors of the degree-$3$
vertex are high-degree vertices.

\vspace{2mm}Let $a,b,c$ and $d$ be four vertices in graph $G$, if
$G$ has four edges $ab$, $bc$, $cd$ and $da$, then we say that
$abcd$ is a \emph{$4$-cycle} in $G$.

\lem{cyclebranch}{Let $abcd$ be a $4$-cycle in graph $G$, then for
any independent set $S$ in $G$, either $a,c\notin S$ or $b,d\notin
S$.}\proof{Since any independent set contains at most $2$ vertices
in a $4$-cycle and the two vertices can not be adjacent, we know
the lemma holds.}

Based on \refl{cyclebranch}, we get the following branching rule.

\vspace{2mm}\noindent \textbf{Branching on a $4$-cycle}\\
\emph{Branching on a $4$-cycle $abcd$ means branching by either
excluding $a$ and $c$ from the independent set or excluding $b$
and $d$ from the independent set.}

\section{A Simple Algorithm}\label{alg}

\begin{figure}[!htbp]
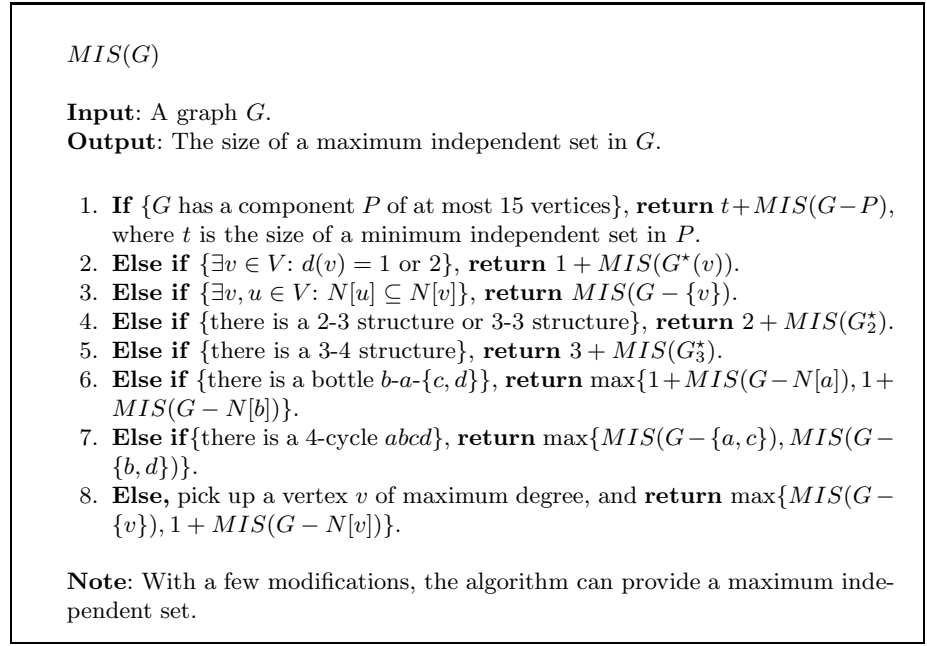
 \setbox4=\vbox{\hsize28pc
\noindent\strut
\begin{quote}
\vspace*{-2mm} \footnotesize {\bf $MIS(G)$} \\

\textbf{Input}: A graph $G$. \\
\textbf{Output}: The size of a maximum independent set in $G$.\\

\begin{enumerate}
\item \textbf{If} \{$G$ has a component $P$ of at most $15$
vertices\}, \textbf{return} $t+MIS(G-P)$, where $t$ is the size of
a minimum independent set in $P$. \item \textbf{Else if}
\{$\exists v\in V$: $d(v)=1$ or $2$\}, \textbf{return}
$1+MIS(G^\star(v))$. \item \textbf{Else if} \{$\exists v,u \in V$:
$N[u]\subseteq N[v]$\}, \textbf{return} $MIS(G-\{v\})$. \item
\textbf{Else if} \{there is a $2$-$3$ structure or $3$-$3$
structure\}, \textbf{return} $2+MIS(G^\star_2)$.\item \textbf{Else
if} \{there is a $3$-$4$ structure\}, \textbf{return}
$3+MIS(G^\star_3)$. \item \textbf{Else if} \{there is a bottle
$b$-$a$-$\{c,d\}$\}, \textbf{return} $\max\{1+MIS(G-N[a]),
1+MIS(G-N[b])\}$. \item \textbf{Else if}\{there is a $4$-cycle
$abcd$\}, \textbf{return} $\max\{MIS(G-\{a, c\}),
MIS(G-\{b,d\})\}$.
 \item \textbf{Else,}
pick up a vertex $v$ of maximum degree, and \textbf{return}
$\max\{MIS(G-\{v\}), 1+MIS(G-N[v])\}$.

\end{enumerate}

\vspace{3mm}\textbf{Note}: With a few modifications, the algorithm
can provide a maximum independent set.
\end{quote} \vspace*{-5mm} \strut}  $$\boxit{\box4}$$ \vspace*{-9mm}
\caption{The Algorithm $MIS(G)$} \label{mis3}
\end{figure}

Our algorithm for the maximum independent set problem is described
in Figure~\ref{mis3}. It works as follows. If the graph has a
component of at most $15$ vertices, we find a maximum independent
set in this component directly (\textbf{Step~$1$}). If the graph
has a degree-$1$ or degree-$2$ vertex, we fold it in
\textbf{Step~$2$}. If the graph has a dominated vertex, we remove
it in \textbf{Step~$3$}. If the graph has a $2$-$3$ structure or
$3$-$3$ structure or $3$-$4$ structure, we fold it in
\textbf{Step~$4$} and \textbf{Step~$5$}. When the graph can not be
reduced, we apply our branching rules. If there is a bottle, we
branch on a bottle (\textbf{Step~$6$}). Else if there is a
$4$-cycle, we branch on a $4$-cycle (\textbf{Step~$7$}). Else in
\textbf{Step~$8$}, we greedily select a vertex of maximum degree
and branch on it by including it in the independent set or
excluding it from the independent set.

 \vspace{-0mm}

 \section{The Analysis}
To analyze the time complexity of our algorithm, we will consider
recurrence relations related to parameter $r$, the number of
degree-$3$ vertices (vertices of degree $\geq 4$ will also be
counted with a weight) in the corresponding graph. When $r=0$, the
graph has only degree-$0$, degree-$1$ and degree-$2$ vertices and
the maximum independent set problem can be solved in linear time.
We use $C(r)$ to denote the worst-case size of the search tree in
our algorithm when the parameter of the graph is $r$. In our
algorithm, it is possible to create a vertex of degree $\geq 4$
when folding. We will regard a degree-$d$ ($d\geq3$) vertex as a
combination of $d-2$ degree-$3$ vertices and count $d-2$ in
parameter $r$. Then when a degree-$d$ vertex is removed, parameter
$r$ will be reduced by $d-2$. When an edge incident on a
degree-$d$ vertex is removed, parameter $r$ will be reduced by
$1$. In the remaining of the paper, when we say a graph has $x$
degree-$3$ vertices, it does not mean that the graph really has
exactly $x$ vertices of degree $3$. In fact, all the vertices of
degree $\geq3$ are counted. Next, we analyze how much $r$ can be
reduced in each step of our algorithm.

\lem{}{After folding a degree-$1$ or degree-$2$ vertex, parameter
$r$ will not increase.}

\lem{folding}{Let $G$ be a graph having no degree-$1$ or
degree-$2$ vertex, then after folding a $2$-$3$ structure or
$3$-$3$ structure or $3$-$4$ structure, or removing a dominated
vertex in $G$, parameter $r$ will be reduced by at least $4$.}
\proof{In each case, a degree-$3$ vertex is removed (or an even
better case occurs), and then we can further reduce $r$ by $3$
from $3$ neighbors of the vertex. Totally $r$ will be reduced by
at least $4$.}

\lem{d1}{Let $G$ be a connected graph. If $G$ has at least $x$
degree-$1$ vertices and $x$ vertices of degree $\geq 3$ (a
degree-$d$ ($d\geq3$) vertex will be regarded as $d-2$ degree-$3$
vertices), then after iteratively folding degree-$1$ vertices
until the graph has no degree-$1$ vertex, parameter $r$ will be
reduced by at least $x$. }

\proof{Let $V'\neq \emptyset$ be the set of vertices of degree
$\geq 2$ in the remaining graph after iteratively folding
degree-$1$ vertices (The lemma obviously holds, when $V'=
\emptyset$). Assume there are $y$ edges between $V''=V-V'$ and
$V'$. After removing $V''$, we can reduce $y$ degree-$3$ vertices
from $V'$. We will prove that there are at least $x-y$ degree-$3$
vertices in $V''$. To prove that, we first construct a new graph
$G'$ from $G$ by contracting $V'$ into a single vertex $v$ and
remove all self-loops incident on it (keeping parallel edges).
Then we only need to prove that except vertex $v$, $G'$ has at
least $x-y$ degree-$3$ vertices.

Since all the $x$ degree-$1$ vertices of $G$ are in $V''$, $G'$
has at least $x'$ degree-$1$ vertices, where $x'=x+1$ when $v$ is
a degree-$1$ vertex and $x'=x$ when $v$ is not a degree-$1$
vertex. Note that a tree with $x'$ degree-$1$ vertices has at
least $x'-2$ degree-$3$ vertices. We know that $G'$ has least
$x'-2$ degree-$3$ vertices ($G'$ is a connected graph). We
consider the following three cases. Case 1: $y=1$. For this case,
$v$ is a degree-$1$ vertex and $x'=x+1$. Then $G'$ has at least
$x'-2=x-1$ degree-$3$ vertices. Case 2: $y=2$. For this case, $v$
is a degree-$2$ vertex and $x'=x$, and $G'$ still has at least
$x-2$ degree-$3$ vertices. Case 3: $y\geq 3$. For this case, $v$
is a degree-$y$ vertex and $x'=x$. Excepting $y-2$ degree-$3$
vertices counted from $v$, there are sill $x-2-(y-2)=x-y$
degree-$3$ vertices.

Therefore, after removing $V''$, $r$ will be reduced by at least
$x$.}

\crl{degree1}{Let $G$ be a graph having not any component of a
path. If $G$ has any degree-$1$ vertex, then we can reduce $r$ by
at least $1$ by iteratively folding degree-$1$ vertices. If $G$
has exactly $2$ degree-$1$ vertices, then we can reduce $r$ by at
least $2$ by iteratively folding degree-$1$ vertices.}

\lem{3db}{Let $G$ be a reduced graph and $v$ a degree-$3$ vertex
in $G$. Then not any degree-$0$ vertex or component of a $1$-path
or component of a $2$-path is created after removing $N[v]$.}
\proof{If a degree-$0$ vertex $u$ is created, then $G$ has a
$2$-$3$ structure $\{v, u\}$-$N(v)$. If a $1$-path $ab$ is
created, then there is a $3$-$3$ structure $\{v, a, b\}$-$N(v)$.
If a $2$-path $abc$ is created, then there is a $3$-$4$ structure
$\{a, c, v\}$-$N(v)\cup\{b\}$.}

\lem{3degree}{Let $G$ be a connected reduced graph of more than
$7$ vertices and $v$ a degree-$3$ vertex in $G$. Then after
removing $N[v]$, parameter $r$ will be reduced by at least $8$.
Furthermore, if each $3$-cycle in $G$ contains at least one vertex
of degree $\geq 4$, then after removing $N[v]$, parameter $r$ will
be reduced by at least $10$.}

\proof{There is at most one edge with both endpoints in $N(v)$,
otherwise $v$ will dominate a neighbor of it. Therefore, there are
at least four edges between $N(v)$ and $N_2(v)$. If
$|N_2(v)|\geq4$, $r$ will be reduced by $4+4=8$ directly after
removing $N[v]$. If $|N_2(v)|\leq3$, it is impossible to create a
component of a $l$-path ($l\geq3$) after removing $N[v]$. By
\refl{d1} and \refc{degree1} and \refl{3db} we know that
eventually $r$ will be reduced by at least $8$.

Next, we assume that in each $3$-cycle in $G$ there is a vertex of
degree $\geq4$. We distinguish the following two cases.
\textbf{Case~$1$}: All vertices in $N(v)$ are degree-$3$ vertices.
In this case, none pair of vertices in $N(v)$ are adjacent and
there are exactly six edges between $N(v)$ and $N_2(v)$, which
means at most $3$ degree-$1$ vertices will be created after
removing $N[v]$. It is impossible to create a component of a path
after removing $N[v]$ (Obviously, no path of length $\geq4$ will
be created. \refl{3db} shows no path of length $\leq2$ will be
created. If a $3$-path is created, then the graph $G$ has only $7$
vertices). So by \refc{degree1}, if a component with $1$ or $2$
degree-$1$ vertices is created after removing $N[v]$, we can
further reduce $r$ by $1$ or $2$ by further reducing degree-$1$
vertices in the component. If a component with $3$ degree-$1$
vertices is created, then the component also contains at least $3$
degree-$3$ vertices, otherwise the only possibility of the
component is that it has $4$ vertices: a degree-$3$ vertex
adjacent with three degree-$1$ vertices, which also implies a
contradiction--the graph $G$ has only $7$ vertices. By \refl{d1},
we still can further reduce $r$ by least $3$. In any case, totally
we can reduce $r$ by at least $4+6=10$.
 \textbf{Case~$2$}: There is a vertex of degree
$\geq4$ in $N(v)$. Then there are at least five edges between
$N(v)$ and $N_2(v)$ (Note that there is at most one edge with both
endpoints in $N(v)$). By \refl{d1} and \refl{3db} we know that $r$
will be reduced by at least $5+5=10$. }

\lem{4degree}{Let $G$ be a connected reduced graph of more than
$8$ vertices and $v$ a vertex of degree $\geq4$ in $G$. Then after
removing $N[v]$, parameter $r$ will be reduced by at least $10$. }

\proof{The lemma obviously holds when $v$ is a vertex of degree
$\geq5$ or a degree-$4$ vertex with $|N_2(v)|\geq 4$. Now we
assume $v$ is a degree-$4$ vertex and $|N_2(v)|\leq 3$.
\textbf{Case~$1$}: $|N_2(v)|=1$. In this case, after removing
$N[v]$, $r$ is reduced by at least $6+4=10$, or $r$ is reduced by
at least $6+3$ and the only vertex in $N_2(v)$ becomes a
degree-$1$ vertex (Note that there are at least $|N(v)|=4$ edges
between $N(v)$ and $N_2(v)$). In the later case, we can reduce $r$
by at least $1$ by folding degree-$1$ vertices.\textbf{ Case~$2$}:
$|N_2(v)|=2$. The two vertices $a, b\in N_2(v)$ are adjacent,
otherwise there is a $3$-$4$ structure $\{v\}\cup N_2(v)$-$N(v)$.
Then after removing $N[v]$, at most one of $a$ and $b$ becomes a
degree-$1$ vertex, otherwise $G$ has only $7$ vertices. Therefore,
we also can reduce $r$ by at least $4$ from $V-N[v]$.
\textbf{Case~$3$}: $|N_2(v)|=3$. If one vertex in $N(v)$ is a
vertex of degree $\geq4$, then the lemma holds. Otherwise, all
vertices in $N(v)$ are degree-$3$ vertices, and then the number of
edges between $N(v)$ and $N_2(v)$ is $4$ or $6$ or $8$. If no
degree-$1$ vertex is created after removing $N[v]$, then $r$ will
be reduced by at least $6+4$ directly (Note that it is impossible
to create two degree-$0$ vertices, and when one degree-$0$ vertex
is created, there are at least three edges between $N(v)$ and
$N_2(v)$ that are incident on the other two vertices in $N_2(v)$).
If some degree-$1$ vertices are created but no path component is
created, then we can further reduce $r$ by at least $1$ by
\refl{d1}. The difficult case occurs when a path component is
created. The path can only be a $1$-path or $2$-path. If it is a
$2$-path, then the graph has only $8$ vertices. Therefore the path
is a $1$-path. If one vertex in the path is a vertex of degree
$\geq4$ in $G$, then after removing $N[v]$, $r$ is reduced by at
least $6+4=10$ directly. If the two vertices in the path are
degree-$3$ vertices in $G$, then there are at least two edges
between $N(v)$ and $N_2(v)$ that are incident on the third vertex
$u$ in $N_2(v)$. So after removing $N[v]$, $r$ is reduced by at
least $6+4=10$, or $r$ is reduced by at least $6+3$ and $u$
becomes a degree-$1$ vertex, folding which will further reduce $r$
by at least $1$. We have checked all the cases and then finished
the proof. }

\lem{bottle}{Let $G$ be a connected reduced graph of more than $7$
vertices.  If $G$ has a bottle, then algorithm $MIS(G)$ will
branch on a bottle with recurrence relation \eqn{8-8_1}{C(r)\leq
2C(r-8),} where $C(r)$ is the worst-case size of the search tree
in our algorithm.

Moreover, if each $3$-cycle in $G$ contains at least one vertex of
degree $\geq4$, then $MIS(G)$  will branch on a bottle with
recurrence relation \eqn{10-10_1}{C(r)\leq 2C(r-10).} }

\proof{Let the bottle called by our algorithm be
$b$-$a$-$\{c,d\}$. Our algorithm will branch by either removing
$N[a]$ or $N[b]$. By \refl{3degree} and \refl{4degree}, we get
\refe{8-8_1} and \refe{10-10_1} directly. }

\lem{4cycle}{Let $G$ be a connected bottle-free reduced graph of
more than $7$ vertices. If $G$ has a $4$-cycle, then algorithm
$MIS(G)$ will branch on a $4$-cycle with recurrence relation
\eqn{8-8}{C(r)\leq 2C(r-8).} Moreover, if each $3$-cycle or
$4$-cycle in $G$ contains at least one vertex of degree $\geq4$,
then $MIS(G)$  will branch on a $4$-cycle with recurrence relation
\eqn{10-10}{C(r)\leq 2C(r-10).}}

\proof{Let the $4$-cycle called by our algorithm be $abcd$. Our
algorithm will branch by removing either $\{a, c\}$ or $\{b, d\}$
from the graph. We look at the branch where $\{a, c\}$ is removed
(It is the same to $\{b, d\}$). Since none of the four vertices is
dominated by others, each of the four vertices will be adjacent to
a vertex different from the four vertices. If after removing $\{a,
c\}$, no degree-$1$ vertex is created, then we can reduce $r$ by
$8$ directly in this branch. If some degree-$1$ vertices are
created, then our algorithm will fold one, say $x$, in the next
step. Obviously, $x$ is a degree-$3$ vertex in the original graph.
The operation of removing $\{a, c\}$ and then folding $x$ is
equivalent to the removing of $N[x]$. We can reduce $r$ by at
least $8$ by \refl{3degree}. Therefore, we get \refe{8-8}.

Next, we prove \refe{10-10}. There is at least one vertex of
degree $\geq4$, say $a$, in the $4$-cycle. We distinguish the
following three cases. \textbf{Case~$1$}: There is only one vertex
of degree $\geq4$ in the $4$-cycle. No matter we remove $\{a, c\}$
or $\{b, d\}$, at least one degree-$1$ will be created. As
discussed above, after further folding a degree-$1$ vertex, we can
reduce $r$ by at least $10$ in each branch by \refl{3degree}. Then
we get \refe{10-10}. \textbf{Case~$2$}: There are exactly two
vertices of degree $\geq4$ in the $4$-cycle and the two vertices
are adjacent to each other in the $4$-cycle (the two vertices are
not $\{a,c\}$ or $\{b,d\}$). Without loss of generality, we can
assume the two vertices of degree $\geq4$ are $a$ and $b$. In the
branch where $\{a, c\}$ is removed, $d$ becomes a degree-$1$
vertex. In the branch where $\{b, d\}$ is removed, $c$ becomes a
degree-$1$ vertex. Then in each branch we will remove $N[v]$ for
some degree-$3$ vertex $v$ in $G$. We still can get \refe{10-10}.
 \textbf{Case~$3$}: There are exactly two
vertices of degree $\geq4$ in the $4$-cycle and the two vertices
are a pair of opposite vertices in the $4$-cycle. Then the two
vertices of degree $\geq4$ are $a$ and $c$ (We have assumed that
$a$ is a vertex of degree $\geq4$). It is easy to see that after
removing $\{b, d\}$, $r$ will be reduced by at least $8$. In the
branch where $\{a, c\}$ is removed, some degree-$1$ vertices are
created (at least $b$ and $d$). Then in this branch we will remove
$N[v]$ for some degree-$3$ vertex $v$ in $G$, where $v$ has two
vertices of degree $\geq4$ $a$ and $c$. Since $G$ has no bottle,
there is not any edge with both endpoints in $N(v)$. Therefore,
there are at least $8$ edges between $N(v)$ and $N_2(v)$. If no
degree-$1$ vertex is created after removing $N[v]$ (\refl{3db}
also shows that no degree-$0$ vertex will be created), then $r$ is
reduced by $6+8=14$ directly. If some degree-$1$ vertices are
created, the case becomes complicated. In fact, as we do in the
proof of \refl{3degree} we can prove that no component of less
than $3$ degree-$3$ vertices will be created. By  \refl{d1}, we
know that $r$ will also be reduced by at least $6+7=13$ in this
branch. We get \eqn{8-13}{C(r)\leq C(r-8)+C(r-13).}
\textbf{Case~$4$}: There are exactly three vertices of degree
$\geq4$ in the $4$-cycle. Without loss of generality, we assume
the remaining degree-$3$ vertex is $c$. After removing $\{a, c\}$,
$r$ will be reduced by at least $10$. After removing $\{b, d\}$,
$r$ will be reduced by at least $12$. We get \eqn{10-12}{C(r)\leq
C(r-10)+C(r-12).} \textbf{Case~$5$}: All the four vertices in the
cycle are vertices of degree $\geq4$. It is clear that $r$ will be
reduced by at least $10$ in each branch. We also get \refe{10-10}.
Since \refe{10-10} covers \refe{8-13} and \refe{10-12}, we know
that the lemma holds. }

\lem{}{Let $G$ be a connected reduced graph of more than $15$
vertices that has no bottle or $4$-cycle. If $G$ has a vertex of
degree $\geq4$, then algorithm $MIS(G)$ will branch on a vertex of
maximum degree with recurrence relation \eqn{6-14}{C(r)\leq
C(r-6)+C(r-14).} }

\proof{Our algorithm will select a vertex $v$ of maximum degree
and branch on it by excluding it from the independent set or
including it in the independent set. In the former branch, $v$ is
removed and $r$ decreases by at least $2+4=6$. In the latter
branch, $N[v]$ is removed. Since $G$ has no bottle or $4$-cycle,
there are at least $8$ vertices in $N_2(v)$. Then in this branch,
$r$ will be reduced by at least $6+8=14$. Therefore, we get
\refe{6-14}.}

\lem{3dv}{Let $G$ be a connected reduced graph of more than $15$
vertices that has no bottle or $4$-cycle. If $G$ is also a
$3$-regular graph, then algorithm $MIS(G)$ can branch with
recurrence relation \eqn{10-14-14}{C(r)\leq C(r-10)+2C(r-14).}}

\proof{Our algorithm will select a degree-$3$ vertex and branch on
it. Since $G$ is $3$-regular graph that has no $3$-cycle or
$4$-cycle, there are exactly $8$ vertices in $N_2(v)$. In the
branch where $N[v]$ is removed, $10$ degree-$3$ vertices are
reduced. So we can branch with recurrence relation
\eqn{10-4}{C(r)\leq C(r-10)+Q(r-4),} where $Q\leq C$ is some
function corresponding to the size of the branch where $v$ is
removed. Next, we focus on refining analysis of $Q$.

In the branch where $v$ is removed, $3$ nonadjacent degree-$2$
vertices are created. Our algorithm will fold the three degree-$2$
vertices in the next step. Let $G'$ be the resulted graph. Then
$G'$ has exactly $3$ degree-$4$ vertices (Note that the original
graph has no $3$-cycle or $4$-cycle. It is impossible to create a
degree-$3$ vertex after folding a degree-$2$ vertex),  and each
$3$-cycle or $4$-cycle in the current graph contains at least one
degree-$4$ vertex. If $G'$ has a bottle or $4$-cycle, we can
branch with $Q(r)\leq 2C(r-10)$ by \refl{bottle} and
\refl{4cycle}. If $G'$ has no bottle or $4$-cycle, we will branch
on a degree-$4$ vertex $v'$. We further distinguish three
different cases. \textbf{Case~$1$}: The other two degree-$4$
vertices are adjacent to $v'$. In this case, we have
$|N_2(v')|\geq 8$ (the three degree-$4$ vertices may form a
triangle). In the branch where $v'$ is removed, $r$ is reduced by
at least $6$, and in the branch where $N[v']$ is removed, $r$ is
reduced by at least $8+8=16$. We get $Q(r)\leq C(r-6)+C(r-16)$.
\textbf{Case~$2$}: There is only one degree-$4$ vertex adjacent to
$v'$. Since there is no bottle and $4$-cycle, we get
$|N_2(v')|\geq9$. In the branch where $N[v']$ is removed, $r$ is
reduced by $7+9=16$. We also get $Q(r)\leq C(r-6)+C(r-16)$.
\textbf{Case~$3$}: There is no degree-$4$ vertex adjacent to $v'$.
We will branch on $v'$ with \refe{6-14} directly, and in the
branch where $v'$ is removed, some other degree-$4$ vertices are
left. We can further branch with \refe{6-14} at least. Then we get
$Q(r)\leq C(r-14)+C(r-6-6)+C(r-6-14)=C(r-12)+C(r-14)+C(r-20)$.

The worst case is that after branching with \refe{10-4} we branch
with \refe{10-10_1} or \refe{10-10}, in which we get $C(r)\leq
C(r-10)+Q(r-4)\leq C(r-10)+2C(r-14)$, as claimed in the lemma. }

Among all the cases in our algorithm, the worst running time
corresponds to recurrence relation \refe{10-14-14}. Since
$C(r)=O(1.0919^r)$ satisfies \refe{10-14-14}, we get

\thm{}{Algorithm $MIS(G)$ can find a minimum independent set in a
degree-$3$ graph in $O^*(1.0919^n)$ time.}

\section{Improvement on $k$-Vertex Cover}
Given a graph $G$ and a parameter $k$, the \emph{$k$-vertex cover}
problem is to decide if $G$ has a vertex cover of size at most
$k$. The $k$-vertex cover problem is one of the most extensively
studied problems in the area of Parameterized Algorithms. In this
section, we show that the $k$-vertex cover problem can be solved
in $O^*(1.1923^k)$ time, which improves
 the previously known result of $O^*(1.1939^k)$ by Chen \emph{et~al.}~\cite{Chen:labeled3vc}.

Nemhauser and Trotter~\cite{Nemhauser:VCkernel} proved the
following theorem:

\prp{}{For a graph $G=(V,E)$ with $n$ vertices and $m$ edges, we
can compute two disjoint vertex sets $C_0,V_0\subset V$ in
$O(\sqrt n m)$ time, such that

$(1)$ Every minimum vertex cover in induced subgraph $G(V_0)$ plus
$C_0$ forms a minimum vertex cover of $G$.

$(2)$ A minimum vertex cover of $G(V_0)$ contains at least $|V_0|/
2$ vertices. }

Our simple algorithm works as follows. Given an instance $(G,k)$
of the $k$-vertex cover problem in degree-$3$ graphs, we first use
Nemhauser and Trotter's algorithm to construct $C_0$ and $V_0$. If
$|V_0|>2k$, then $G$ does not have a vertex cover of size at most
$k$. Else we use our algorithm presented in Section~\ref{alg} to
find a maximum independent set $S$ in $G(V_0)$ in
$O^*(1.0919^{|V_0|})=O^*(1.1923^{k})$ time. Then $C_0+V_0-S$ is a
minimum vertex cover of $G$. If $k>|C_0+V_0-S|$, then $G$ does not
have a vertex cover of size at most $k$. Else $C_0+V_0-S$ is
satisfied vertex cover.

\thm{}{The $k$-vertex cover problem in degree-$3$ graphs can be
solved in $O^*(1.1923^k)$ time.}

\section{Concluding Remarks}\label{conclusion}
In this paper, we have presented a simple $O^*(1.0919^n)$-time
algorithm for the minimum independent set problem in degree-$3$
graphs and a simple $O^*(1.1923^k)$-time algorithm for the
$k$-vertex cover problem in degree-$3$ graphs. Both algorithms
improve previously known algorithms.

Unlike most previous algorithms, our algorithms do not contain
many branching rules. We use two new branching techniques, called
branching on a bottle and branching on a $4$-cycle, to avoid
tedious examinations of the local structures. In fact, new
branching rules catch the structural properties of small cycles in
graphs, which make our algorithms simple and practical. It is easy
to see that many previous algorithms can apply these two new
branching rules to simplify the description and analysis.

Our algorithm for the maximum independent set problem is analyzed
by measuring the number of degree-$3$ vertices. We have checked
that our algorithm $MIS(G)$ can also be analyzed by measuring
parameter $m-n+t$ to get the same running time bound, where $m$ is
the number of edges, $n$ the number of vertices, and $t$ the
number of tree components in the graph. Readers may note that the
algorithms presented by F{\"u}rer~\cite{Furer:ISsparse} and
Bourgeois \emph{et al.}~\cite{Bourgeois:3IS} are analyzed by
measuring $m-n$. In fact, their algorithms also need to consider
the tree components created in the graphs and they have a separate
section to analyze them. We guess that considering the tree
components in the parameter may lead to a clearer analysis.

\bibliographystyle{splncs}

\end{document}